\documentclass[twocolumn]{aastex62}

\newcommand{\beq}	{\begin{equation}}
\newcommand{\eeq}	{\end{equation}}
\newcommand{\beqa}{\begin{eqnarray}}
\newcommand{\eeqa}{\end{eqnarray}}

\newcommand{\e}	{$^{-1}$}
\newcommand{\msun}	{M$_\odot$}
\newcommand{\rsun}	{R$_\odot$}

\newcommand{\lx}	{L$_{\rm X}$}

\newcommand{\ee}	{$^{-2}$}
\newcommand{\eee}	{$^{-3}$}
\def\simlt{\lower.5ex\hbox{$\; \buildrel < \over \sim \;$}}
\def\simgt{\lower.5ex\hbox{$\; \buildrel > \over \sim \;$}}

\graphicspath{{./}{figures/}}

\shorttitle{Cosmic-Ray Mediated Disks}
\shortauthors{Offner et al.}

\usepackage{natbib}
\begin{document}

\title{Impact of Cosmic-Ray Feedback on Accretion and Chemistry in Circumstellar Disks} 

\correspondingauthor{Stella Offner}
\email{soffner@astro.as.utexas.edu}

\author{Stella S. R. Offner}
\affil{Department of Astronomy, The University of Texas at Austin, 2500 Speedway, Austin, TX 78712, USA}
\author{Brandt A. L. Gaches}
\affil{Department of Astronomy, University of Massachusetts, 710 N Pleasant St, Amherst, MA 01003, USA }
\affiliation{I.~Physikalisches Institut, Universit\"at zu K\"oln, Z\"ulpicher Stra{\ss}e 77, Germany}
\author{Jonathan R. Holdship}
\affil{Department of Physics and Astronomy, University College London, Gower Street, WC1E 6BT, London, UK}



\begin{abstract}
We use the gas-grain chemistry code {\sc uclchem} to explore the impact of cosmic-ray feedback on the chemistry of circumstellar disks. We model the attenuation and energy losses of the cosmic-rays as they propagate outwards from the star and also consider ionization due to stellar radiation and radionuclides. For accretion rates typical of young stars $\dot M_* \sim 10^{-9}-10^{-6}$\,\msun yr\e, we show that cosmic rays accelerated by the stellar accretion shock produce a cosmic-ray ionization rate at the disk surface $\zeta \gtrsim 10^{-15}$ s\e, at least an order of magnitude higher than the ionization rate associated with the Galactic cosmic-ray background. The incident cosmic-ray flux enhances the disk ionization at intermediate to high surface densities ($\Sigma > 10$ g cm\ee) particularly within 10 au of the star. We find the dominant ions are C$^+$, S$^+$ and Mg$^+$ in the disk surface layers, while the H$_3^+$ ion dominates at surface densities above 1.0\,g cm\ee. We predict the radii and column densities at which the magneto-rotational instability (MRI) is active in T Tauri disks and show that ionization by cosmic-ray feedback extends the MRI-active region towards the disk mid-plane. However, the MRI is only active at the mid-plane of a minimum mass solar nebula disk if cosmic-rays propagate diffusively ($\zeta \propto r^{-1}$) away from the star.  The relationship between accretion, which accelerates cosmic rays, the dense accretion columns, which attenuate cosmic rays, and the MRI, which facilitates accretion, create a cosmic-ray feedback loop that mediates accretion and may produce luminosity variability.
\end{abstract}


\keywords{accretion, accretion disks –- magnetic fields –-  astrochemistry -- stars: low-mass –- cosmic rays.}


\section{Introduction} \label{sec:intro}

The local degree of gas ionization shapes many fundamental processes in star formation. It dictates which chemical reactions dominate and, consequently, the abundances of various atoms and molecules \citep{wakelam12}. These in turn drive heating and cooling processes, which determine the gas temperature. Ionization influences gas dynamics by determining the strength of the coupling between the gas and the magnetic field, which sets the importance of non-ideal magnetic effects like ambipolar diffusion \citep{fiedler93}.  Ionization is especially critical to the properties and evolution of circumstellar disks \citep{dutrey14}. The level of ionization determines whether the magneto-rotational instability (MRI) is active and thereby dictates the efficiency of angular momentum transport and accretion \citep{balbus91,gammie96}. Regions with low ionization near the disk midplane, where the MRI is inactive, provide conditions favorable for planetesimal formation \citep{gressel12}. At late times, photo-ionization facilitates disk dispersal \citep{alexander14}. 

Ionization in circumstellar disks is produced by a variety of processes. Far-ultaviolet (FUV) radiation, defined as photons with energies between $\sim6-13.6$ eV, from the host star, other nearby stars or the diffuse Galactic background ionize the disk surface layers \citep{semenov04,perez11}. Stellar X-ray photons, predominantly produced by active stellar coronae, have a longer mean-free-path and ionize gas deeper within the disk \citep{semenov04,glassgold07}. Short and long-lived radionuclides within the disk ionize through their decay products: $\sim1$ MeV photons, positrons, electrons and $\alpha$-particles \citep{cleeves13b,padovani2018}. Ionization due to radionuclides only dominates near the midplane, however,  where other sources are absent. High-energy protons and electrons, i.e., cosmic-rays (CRs), which are able to reach gas with surface densities $>100$ g cm\ee~ may also provide a substantial source of ionization \citep{UN81,padovani2018}. Although the Galactic CR background is significant ($\zeta\sim 10^{-16}$\,s\e, \citealt{indriolo2012}), most of these CRs are likely screened by the magnetosphere of the host star \citep{cleeves13a}. However, CRs coming from the star itself could provide significant ionization to the inner disk \citep{rodgers17}. Recent models suggest the CR flux may be enhanced near accreting sources by CRs accelerated within jets or accretion shocks \citep{padovani2016,gaches18b}. 

Measuring the CR flux directly is challenging, since our Sun's magnetosphere screens low-energy CRs. The best measurements of the interstellar cosmic ray background have been recorded by Voyager I, which has reached interstellar space; however, the shape of the spectrum of the Galactic CR background remains uncertain \citep[e.g.,][]{padovani2018}. The CR ionization rate is consequently measured indirectly through ions, such as H$_3^+$, which are produced through CR reactions with neutral gas \citep{indriolo2007,indriolo2015}. These measurements return the CR ionization rate for different pointings and suggest the ionization rate due to the CR background is $\sim 10^{-16}$\,s\e~on average \citep{indriolo2012}. Observations towards protostellar sources, such as OMC-2 FIR4, suggest the CR ionization rate may be higher near young sources \citep{ceccarelli2014,favre2018}. Astrochemical models can explain this ionization provided CRs are accelerated locally by stellar accretion shocks \citep{gaches18b,gaches19}. The measured and predicted ionization rates are sufficiently high to suggest such CRs significantly will also affect the disk chemistry, properties and evolution.

In this work we explore the impact of CRs accelerated by the accreting young host star on its circumstellar disk. We focus on slightly older, T Tauri-type sources, which host protoplanetary disks, have non-negligible accretion ($\dot M_* \gtrsim 10^{-9}$\,\msun\,yr\e) but are no longer embedded in a dense envelope. We consider the impact of different accretion rates, gas-magnetic field coupling and CR propagation assumptions. In \S\ref{parameters} we describe our model setup and parameter choices. In \S\ref{results} we present the resulting ionization rates, disk chemistry and associated impact on the MRI-active regions. We discuss our results in \S\ref{discussion}, including the implications for observations and comparison to prior work. We summarize our conclusions in \S\ref{conclusions}.

\section{Study Parameters}\label{parameters}

\subsection{Disk Initial Conditions and Chemistry}

We adopt the time-dependent, gas-grain chemistry code {\sc uclchem} to compute the abundance distribution \citep{holdship17}. {\sc uclchem} treats each zone independently, taking as inputs the initial species abundance, local temperature, dust extinction to the surface, local far ultra-violet (FUV) radiation field, cosmic ray (CR) ionization rate and number density. We perform the calculations by separately modelling the vertical disk structure at each radius. Thus, we assume radiation and CRs enter the disk from the surface, and we neglect contributions from radiation and CRs that may have propagated horizontally through the disk. 

We define the disk density distribution in terms of the total column density, $N_{\rm tot} = N_{\rm H} + 2N_{{\rm H}_2}$,  where the surface density $\Sigma \equiv N_{\rm tot}m_{\rm H}$ and $m_{\rm H} =1.67 \times 10^{-24}$~g. The local number density of hydrogen is $n_{\rm tot} \sim N_{\rm tot}/h$. Given a flared disk, for a given column density, the radiation is incident at a grazing angle $\theta$, such that the effective column of extinction is $\sim N_{\rm tot}/\theta$. Following \citet{perez11} we adopt $\theta = 0.3$. We assume the inner disk is truncated at the magnetosonic radius \citep[e.g.,][]{manara14,hartmann2016}: 
\begin{equation}
    r_m = \left(\frac{3B_*^2 R_*^6}{2 \dot M_*(G M_*)^{0.5}}\right)^{2/7},
\end{equation}
where $M_*$ is the star mass, $\dot M_*$ is the accretion rate, $R_*$ is the stellar radius, and $B_*$ is the stellar magnetic field.
for our fiducial model,  $r_{\rm in}=0.021$ au. We adopt $r_{\rm out} = 300$ au for outer disk radius. 

{\sc uclchem} does not solve the equations of thermal balance, so we assume the disk is vertically isothermal and set the gas temperature according to \citep{glassgold07,perez11}:  
\begin{equation}
T = 384 r_{\rm au}^{-3/7}~{\rm K},
\end{equation}
where $r_{\rm au}$ is the disk radius in au. 
This yields similar temperature values to \citep{offner09}:
\begin{equation}
T = \left( \frac{L_*}{4 \pi \sigma r^2} \right)^{1/4} 
=  395 L_0^{1/4} r_{\rm au}^{-1/2}~{\rm K},
\end{equation}
assuming a 1$L_\odot$ star
where  $L_0$ is the luminosity in $L_{\odot}$.

We adopt the chemical network from \citet{quenard18}, which is constructed using the UMIST database plus additional reactions to hydrogenate C, N, and O atoms \citep{mcelroy2013,wakelam2013}. 
The network contains 522 species, including 211 on grain surfaces, and computes 3217 reactions altogether.
We also consider a reduced network, in which only reactions with hydrogen, helium, carbon, oxygen and sulfur are considered.
Table \ref{tab:abund} lists the elemental abundances for each network. We initialize the chemistry such that half of the hydrogen is molecular and half atomic, all of the carbon and silicon are ionized and the remaining elements are neutral. 

The formation and freeze-out of species on dust grains is a function of the dust number density and number of binding sites on the grain surface \citep[see Appendix A in][]{quenard18}. Here we assume a gas-to-dust mass ratio of 100, a dust grain mass density of $\rho=3$ g cm\eee~ and a uniform grain size population.   Significant grain growth is expected in protoplanetary disks relative to the interstellar medium (ISM), especially near the disk mid-plane \citep[e.g.,][]{dullemond04,dalessio06}. As the grain size increases, the total grain surface area declines as $a_{\rm gr}^{-1}$ (assuming a fixed gas-to-dust mass ratio). This reduces the area available to capture charged particles, which increases the gas ionization fraction, thereby influencing the diffusivity \citep{wardle07}. 
Consequently, we expect that the details of the MRI are sensitive to the grain distribution.
To quantify the effect of grain growth, we investigate three dust sizes. We adopt a grain radius of $a_{\rm gr}=1.0\,\mu$m for the fiducial value and compare this with small dust typical of the ISM ($a_{\rm gr}=0.1\,\mu$m) and large dust representative of significant dust coagulation ( $a_{\rm gr=}10\,\mu$m). In practice, we expect the dust grain distribution to vary with scale height. However, a more sophisticated grain treatment than we apply here is required to capture this effect.

We make three modifications to {\sc uclchem} to include physical processes relevant for circumstellar disks. First, we modify the treatment of the CR ionization rate to permit variation as a function of position. This allows us to model the CR attenuation within the disk (see \S\ref{sec:cr}). 
The CR ionization rate is updated after each chemical solve, which also allows the CR flux to vary as a function of time.

Second, we include ionization due to X-rays. We compute the extinction of the X-ray radiation field using the column density to the disk surface, i.e., similar to the treatment of the FUV field (see \S\ref{sec:xray}). 
Last, we include ionization due to short-lived radionuclides. We treat this radionuclide ionization rate as a constant throughout the disk. Since radionuclides generally produce much lower ionization compared to other sources in our problem it acts as an ionization floor. We add the X-ray and radionuclide ionization rates to the CR-ionization rate at each location to account for all secondary photons produced through X-ray and CR ionizations as well as non-thermal desorption reactions, whereby molecules are ejected from grain surfaces \citep{holdship17}.
{\sc uclchem} also accounts for secondary electrons produced through CR interactions but assumes they are thermalized, so they are not treated as new CR particles. 
We describe each source of ionization and the associated parameters in detail in \S\ref{sec:ionrate}.

\begin{deluxetable}{l|ccc}
\tablecaption{Elemental Abundances \tablenotemark{a} \label{tab:abund}}
\tablehead{ \colhead{Species} & \colhead{Reduced} &  \colhead{Full} }

\startdata
He & $0.1$ & $0.1$  \\
C$^+$\tablenotemark{a} & $10^{-4}$ & $10^{-4}$ \\
O\tablenotemark{a} & $10^{-4}$ & $10^{-4}$  \\
S \tablenotemark{a} & $10^{-5}$ &  $10^{-5}$  \\
N \tablenotemark{a} & 0 & $6.1 \times 10^{-5}$  \\
Mg\tablenotemark{a} & 0 & $3.8 \times 10^{-5}$  \\
Si$^+$\tablenotemark{a} & 0 & $10^{-7}$ \\
\enddata
\tablenotetext{a}{Abundance relative to to the total nuclear hydrogen density.}
\tablecomments{We compare two networks: (1) a simple network with only hydrogen, helium, carbon, oxygen and sulphur reactions and (2) a network also including magnesium, nitrogen and silicon reactions \citep{quenard18}.}
\end{deluxetable}

\subsection{Stellar Properties}\label{sec:stellarprop}

We assume the host star has 1~\msun~ and explore several different accretion rates as summarized in Table \ref{tab:stars}. 
The sources represent T Tauri stars, which have approximately reached their final mass, but continue to accrete via magnetospheric accretion from their disks at lower rates. T Tauri stars are observed to have magnetic fields of $\sim 1$\,kG, so we adopt that as our fiducial stellar field \citep{johns-krull07}. 

We define the parameter, $f_{\rm acc}$, which represents the covering fraction of accretion hot spots on the stellar surface. 
The covering fraction appears to correlate with the accretion rate, i.e., stars experiencing lower rates of accretion have lower covering fractions \cite[e.g.,][]{hartmann2016}.  Here, we assume the covering fraction is proportional to the accretion rate:  $f_{\rm acc} = 10^{-3}-10^{-1}$ for accretion rates of $\dot M_* = 10^{-9}-10^{-6}$\,\msun\,yr\e.

\begin{deluxetable*}{l|cccccccc}
\tablecaption{Model Properties \tablenotemark{a} \label{tab:stars}}
\tablehead{ 
\colhead{Description}  & \colhead{$\dot M_*$~(\msun yr\e)} & \colhead{$R_*$~(\rsun)}&  \colhead{$r_A$~(au)} &  \colhead{$f_{\rm acc}$} &  \colhead{$\epsilon$} &  \colhead{\lx~(erg s\e)} & \colhead{$a$} & \colhead{$a_{\rm gr}$($\mu$m)} }
\startdata
Fiducial Model  & $10^{-7}$ & 1.5 & 0.021 & 0.01 & 1.0 & $5 \times 10^{29}$ & 2  & 1.0  \\ 
High Accretion   & $10^{-6}$ & 1.5 & 0.007 & 0.1 & 1.0  & $5 \times 10^{29}$  & 2 & 1.0 \\
Low X-ray Luminosity  & $10^{-7}$ & 1.5 & 0.021 & 0.01 & 1.0 & $1 \times 10^{29}$ & 2 & 1.0 \\
Partial Coupling  & $10^{-7}$& 1.5  & 0.021 & 0.01 & 0.1 & $5 \times 10^{29}$ &  2 & 1.0 \\
Low Coupling  & $10^{-7}$ & 1.5 & 0.021 & 0.01 & 0.01 & $5 \times 10^{29}$ &  2  & 1.0 \\
Low Accretion  & $10^{-9}$ & 1.0 & 0.039 & 0.001 & 1.0 & $5 \times 10^{29}$  & 2 & 1.0  \\ 
Small Dust  & $10^{-7}$ & 1.5 & 0.021 & 0.01 & 1.0 & $5 \times 10^{29}$ & 2  & 0.1  \\ 
Large Dust  & $10^{-7}$ & 1.5 & 0.021 & 0.01 & 1.0 & $5 \times 10^{29}$ & 2  & 10.0  \\
Diffusive Attenuation  & $10^{-7}$ & 1.5 & 0.021 & 0.01 & 1.0 & $5 \times 10^{29}$ & 1  & 1.0  \\ 
Low Accretion, Diffusive Attenuation  & $10^{-9}$ & 1.0 & 0.039 & 0.001 & 1.0 & $5 \times 10^{29}$  &  1 & 1.0   \\ 
\enddata
\tablenotetext{a}{Accretion rate, stellar radius, Alfv\'en radius (disk inner radius), covering fraction of accretion, coupling between CRs and magnetic field between the star and disk,  X-ray luminosity, attenuation exponent (CR $\propto r^{-a}$) and dust grain radius.  All runs assume $L_{FUV} = 10^{30}$ erg s\e and a stellar field of $B=1$ kG.}
\tablecomments{ Cosmic-rays are accelerated assuming diffusive shock acceleration from the accretion shock at the stellar surface and attenuated as described above. See \citet{gaches2018b} for details. }
\end{deluxetable*}

\subsection{Shock Properties and Cosmic Ray Acceleration}\label{sec:crprop}

In order for diffusive shock acceleration to occur, the flow must satisfy two properties. First, the gas must be supersonic, i.e., the shock velocity must exceed the thermal sound speed, $v_s>c_s$, and second, the shock must be super-Alfv\'enic, $v_s > v_A$, where $v_A = B/\sqrt{4 \pi \rho_i}$, $B$ is the magnetic field and $\rho_i$ is the ion number density. The shock density can be written in terms of the accretion rate, $\dot M_*$, stellar radius, $R_*$, and accretion covering fraction: 
\begin{equation}
\rho_i = \frac{\dot M_*}{f_{\rm acc} 4 \pi R_*^2 v_s m_{\rm H} },    
\end{equation}
where $\mu=0.6$ for an ionized gas.
We set the shock velocity of the freefall velocity at the stellar surface, $v_s = \sqrt{2GM_*/R_*}$.
Combining these expressions we can write the Alfv\'en criterion in terms of the stellar parameters and accretion rate
\begin{eqnarray}
f_A &\equiv& \frac{v_s}{v_A} \nonumber \\ 
&=& 3.7 \left( \frac{M_*}{1\,M_\odot}\right)^{1/4}\left( \frac{ \dot M_*}{ 10^{-7} \, M_\odot \,{\rm yr}^{-1}} \right)^{1/2}   \left( \frac{B}{1\,{\rm kG}}\right)^{-1} \times  \nonumber \\ 
&& \left( \frac{f_{\rm acc} }{0.01}\right)^{-1/2}  \left( \frac{R_* }{R_\odot}\right)^{-5/4}.    \label{eq:alfven}
\end{eqnarray}

This expression shows that the condition tends to be met for higher accretion rates and higher-mass stars, while ``puffier" radii and lower accretion rates tend to reduce $v_s$, limiting the acceleration of cosmic rays.

For the initial CR spectrum, we adopt the calculation of \citet{gaches2018b} for CRs accelerated through an accretion shock. This produces particles with energies up to a few GeV and has an energy dependence $\propto E^{-2}$. The particles loose energy through a variety of mechanisms as they propagate away from the source. We discuss our treatment of the attenuation of the CR spectrum and calculation of the CR ionization rate in \S \ref{sec:cr}.

\subsection{Ionization Rates}\label{sec:ionrate}

\begin{figure*}[ht!]
\plotone{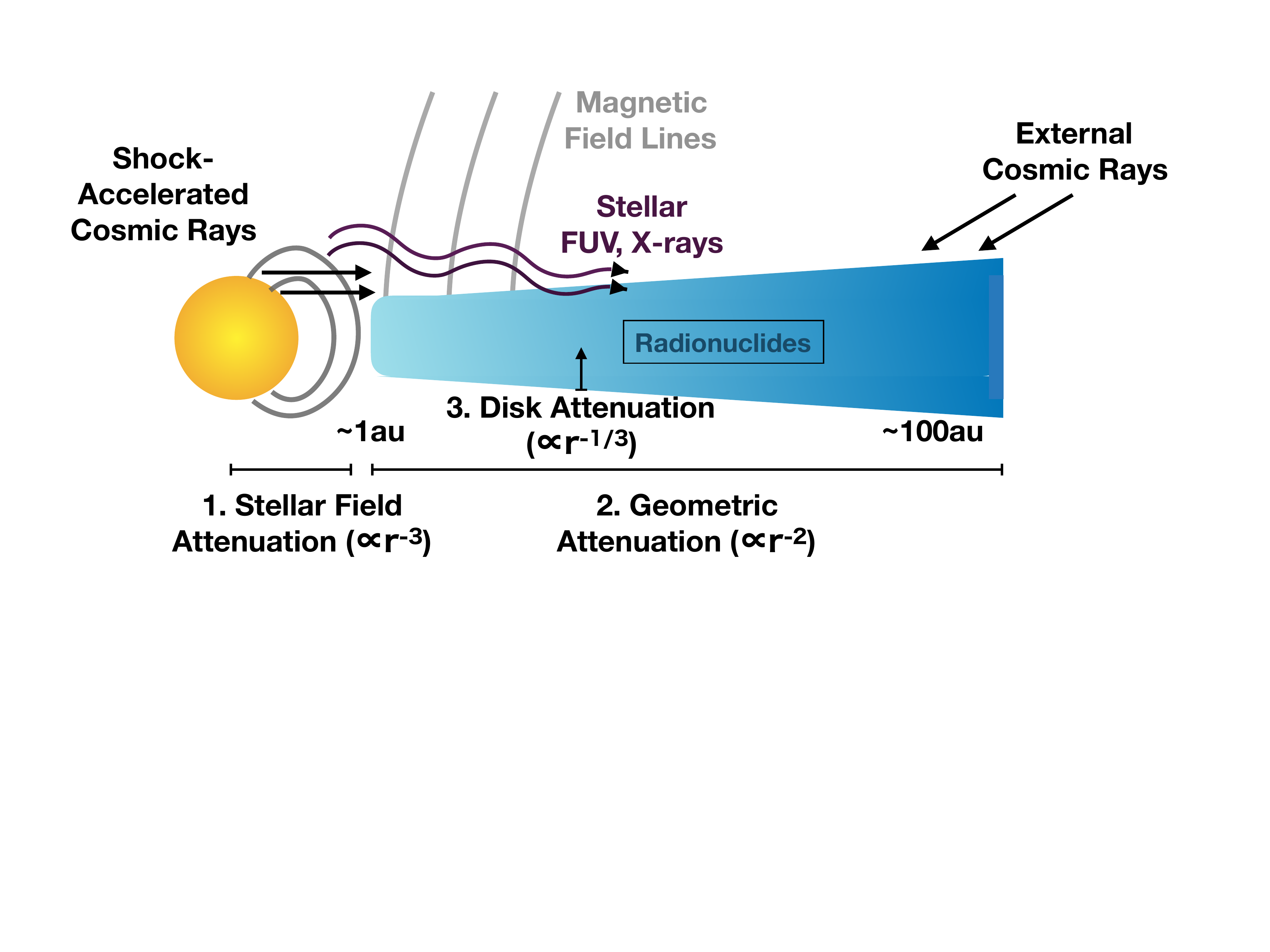}
\vspace{-1.7in}
\caption{Schematic of our model.  \label{fig:schematic}}
\end{figure*}

Circumstellar disks have four primary sources of ionization: X-rays, cosmic rays, FUV radiation and radionuclides. Each of these has a different relationship with radius and surface density and thus serves as the primary source of ionization in different parts of the $r-\Sigma$ parameter space. Figure \ref{fig:schematic} shows a schematic of the sources of ionization and the associated regions. Here we consider only local sources of ionization, namely due to the central stellar source, which generally dominate over external sources such as the Galactic X-ray, CR and FUV backgrounds.

\subsubsection{Cosmic Rays}\label{sec:cr}

CRs can penetrate column densities up to $\Sigma \sim 100$\,g\,cm\ee~before attenuating significantly, potentially allowing them to reach the disk midplane \citep{UN81}. In practice, however, the influence of external CRs may be negligible due to shielding by stellar magnetosphere or ``T-Tauriosphere" \citep{cleeves13a}. In contrast, CRs from the source itself do not have this limitation \citep{rodgers17}. In particular, CRs accelerated at the stellar surface due to accretion shocks may provide a significant source of ionization \citep{padovani2009,gaches2018b,padovani2018}. \citet{gaches2018b} showed that accreting, protostars are a strong source of CRs, which impact the ionization of the dense core and larger molecular cloud. Here, we explore the impact of CRs accelerated in accretion shocks on the chemistry and dynamics of circumstellar disks.

Like photons, CRs attenuate geometrically as they propagate outward from the central star and also suffer energy losses via interactions with gas along their path. Exactly how CRs propagate outwards is not known, since it is a function of the magnetic field geometry, gas density and degree of turbulence. Qualitatively, the flux of CRs declines as $\propto r^{-a}$, where the two limits are free-streaming  ($a=2$) and diffusive ($a=1$). \citet{gaches2018b} suggested that the high CR fluxes accelerated by the accretion shock require attenuation $\propto r^{-2}$ or steeper to preserve the observed cold gas temperatures of dense cores. This work, however, did not consider  attenuation by the accretion flow near the star, which may produce significant attenuation, and which we investigate here. For our fiducial model we adopt $a=2$ between the inner and outer disk radius, but we also consider the case where $a=1$. 

There are two additional competing effects that modulate the CR flux. First, mirroring, i.e., the reflection of CRs due to the convergence of field lines, can significantly reduce the flux of cosmic rays \citep{padovani11}. This acts to reduce the CR flux impinging on circumstellar disks from external sources \citep{cleeves13a}. In contrast, converging field lines can also focus the CR flux via funneling \citep{padovani11,silsbee18}. In the case of CRs accelerated at the stellar surface, the field lines {\it diverge}, therefore, the CR flux effectively undergoes inverse funneling. 

\begin{figure*}[ht]
\epsscale{1.15}
\plottwo{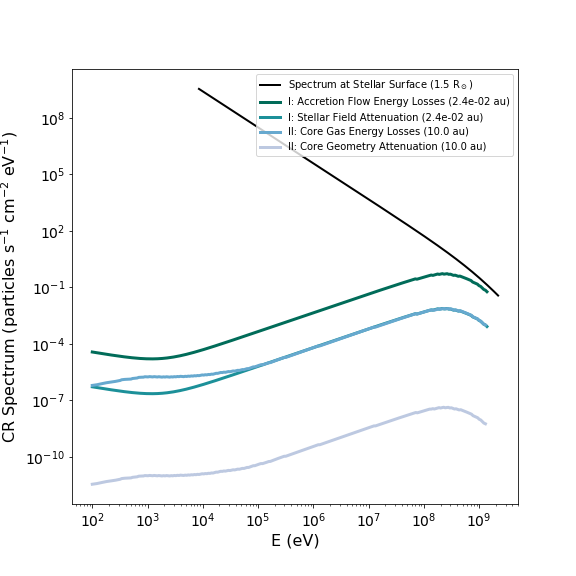}{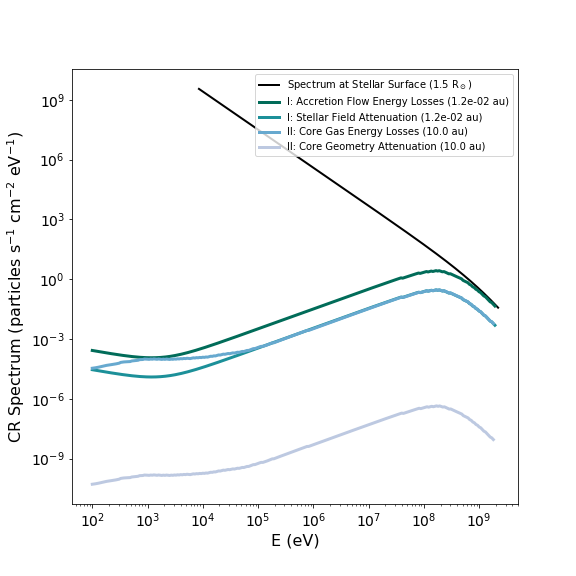}
\caption{Cosmic-ray spectrum at different regions for the fiducial model (left:  $m=1$\msun, $r=1.5$\rsun, $\dot M_* = 10^{-7}$\msun yr\e and $f_{\rm acc} = 0.01$) and the high-accretion model (right: $m=1$\msun, $r=1.5$\rsun, $\dot M_* = 10^{-6}$\msun yr\e and $f_{\rm acc} = 0.1$). } \label{fig:crspec}
\end{figure*}

To determine the local CRIR, we begin with the initial shock-accelerated spectrum of CRs at the stellar surface and propagate it, including geometric attenuation and energy losses, through three regimes: (I) between the star and the inner edge of the accretion disk, (II)  between the radius of the inner disk and the disk surface and (III) inside the disk (see Figure \ref{fig:schematic}).  Figure \ref{fig:crspec} shows the evolution of the CR spectrum as it evolves with radius and is shaped by each of the different processes we describe below.

In region I, the CR spectrum is attenuated via inverse funneling and undergoes energy losses due to interactions with gas upstream of the shock. 
To compute the energy losses due to propagation through accreting material, we assume a completely ionized gas and compute the loss function as a combination of Coulomb  and pion-production losses \citep{mannheim1994,schlickeiser2002}. Figure \ref{fig:crspec} shows that losses due to the accretion flow significantly reduces the CR flux, an impact which is particularly acute for lower energy particles.

In region I, the flux of CRs is also diminished due to inverse funneling. We define the funneling attenuation factor:
\begin{equation}
   f_{\rm funnel} = \frac{B_*}{B_{\rm disk}}, 
\end{equation}
where $B_*$ is the magnetic field at the stellar surface and $B_{\rm disk}$ is the field at the disk surface \citep{cleeves13a}. Young stars have surface magnetic fields of $\sim 1$ kG, and their field structure can be described as the sum of a series of multi-pole moments \citep{romanova15}.  \citet{donati08} find that the poloidal field of T Tauri star BP Tauri is largely comprised of dipole (50\%) and octupole (30\%) components with a small fraction in higher order components (10\%). The approximately 10\% field remaining is in a toroidal component.  Therefore, we adopt
\beq 
f_{\rm funnel} \propto 0.5 r^{-3} + 0.3 r^{-7} + 0.1 r^{-9}.
\eeq 
This implies that if CRs are well-coupled to the field, the spectrum declines more steeply than the often-assumed $r^{-2}$ free-streaming limit. For our parameters, inverse funneling typically reduces the flux by $\sim 10-100$ between the star and inner disk.

In region II, we assume the field attenuates geometrically (e.g., $a=2$) and undergoes small energy losses due to interactions with neutral material above the disk ($N_{\rm H} < 10^{20}$ cm \ee). 
To compute the energy losses due to gas above the disk we adopt the loss function from \citet{padovani2009}, which assumes the gas is cold and mostly molecular, and compute the column density assuming a number density of $n_{\rm H} = 10^4$ cm\eee:
\begin{equation}
N_{\rm H,Core}(r) = \int_{r_A}^{r} n_{\rm H} ds.    
\end{equation}
Interactions between the highest energy CRs and the neutral gas produce new lower energy CRs, which increase the number of CRs with energies $E\sim 10^2-10^5$ eV as shown in Figure \ref{fig:crspec}. 
We then compute the effective ionization rate at the disk surface by integrating over the spectrum. 

\begin{figure}[ht!]
\epsscale{1.2}
\plotone{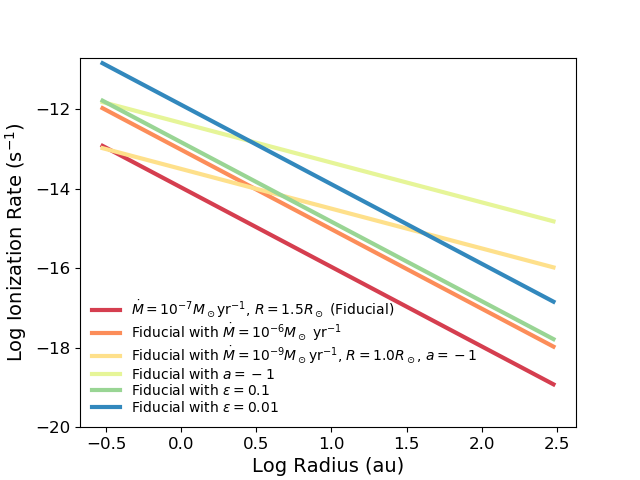}
\caption{Ionization rate as a function of radius at the disk surface for various model parameters.} \label{fig:crion}
\end{figure}

Figure \ref{fig:crion} shows the surface ionization rates as a function of radius for various models listed in Table \ref{tab:stars}. At most radii, the ionization rates are well-above the Galactic cosmic-ray background ionization rate of $\zeta \sim 10^{-16}$s\e \citep{indriolo2007,indriolo2012}. All models predict cosmic-ray ionization rates above the ionization rate, $\zeta \leq 10^{-20}$\,s\e, if Galactic CRs are efficiently shielded by the T-Tauriosphere \citep{cleeves13a}.

Within the disk (region III) we adopt the expression from \citet{padovani2018} to estimate the decline in ionization rate with disk column density, $N_{\rm H}$: 
\beq
\zeta_{CR}= \zeta_0 \left( \frac{N_H}{10^{18}\, {\rm cm}^2}\right)^{-0.34}, \label{diskion}
\eeq 
where  $\zeta_0$ is the ionization rate at the surface of the disk. This formulation takes into account the effect of a wider range secondary and tertiary products. The dominant source of ionization at surface densities greater than 100 g cm$^{-2}$ is electron-positron production as a result of pion production. We note that the details of the attenuation are sensitive to the spectrum shape, treatment of the interactions, and the inclusion of secondary particles, i.e., additional high energy particles produced by CR interactions.   \citet{padovani2018} adopts a CR spectrum representing the CR background of the interstellar medium, which follows a different energy dependence than our CR spectrum at the disk surface. However, the results of \citet{padovani2018} suggest that the ionization rate declines according to a weak power law when $N_H < 10^{25}$ cm\ee~ for both ``high" and ``low" CR spectra, so we assume the attenuation will be qualitatively similar in our case. 

One additional free parameter is the amount of turbulence in the magnetic field. In principle, the CRs and field are well-coupled, since the Larmor radius is small: a proton with momentum 3 Gev/c in a 1 kG field has $r_L < 1.4 \times 10^{-7}$ \rsun. The particles would not be expected to scatter more than once per $r_L$. However, turbulence in the field can change the rate of diffusion
\citep{schlickeiser2002}. Dense cores are expected to be at least sub-sonically turbulent, while outflows and disk winds may drive additional turbulence \citep{offner17}. How this manifests as magnetic turbulence on au scales, however, is poorly constrained. \citet{rodgers17} define a turbulent diffusion parameter, $D$, to explore variation in CR propagation, which they vary between $3-380 r_L$. 
They show that the attenuation of the CR spectrum varies between $\propto r^{-2}$ (free-streaming) ($D = 3 r_L$) and $\propto r^{-1}$  ($D = 380 r_L$) depending on the assumed turbulence and number of scatterings.
For the purposes of our fiducial parameters we adopt the attenuation factors as described above and define a coupling parameter, $\epsilon$, which we explore in \S\ref{cratten}.


\subsubsection{Far Ultra-Violet Radiation}\label{sec:fuv}

FUV radiation from the central star is significant at the disk surface, but it rapidly attenuates with increasing column density. Its contribution becomes negligible by $\Sigma \sim 0.01 $g cm\ee~ \citep[e.g.,][]{perez11}. Thus, ionization due to FUV irradiation is confined to the disk surface layers. 

We adopt a fiducial value for the FUV luminosity of $L_{\rm FUV} = 10^{30}$ erg s\e.  We then compute the local radiation field as a function of radius:
\beqa
G_0 &=& \frac{L_{\rm FUV}}{4 \pi r^2}\left(\frac{1}{1.6\times 10^{-3}\, {\rm erg\, s}^{-1}\,{\rm cm}^{-2}} \right) \\ 
&=& 1.4 \times 10^4 \left( \frac{L_{\rm FUV}}{10^{30}\, {\rm erg s}^{-1}} \right) \left(\frac{r}{3\, {\rm au}} \right)^2 {\rm ~Habing}.
\eeqa
The ionization rate due to FUV photons is then:
\beq
\zeta_{\rm FUV} = G_0 a_X {\rm exp}(-b_X A_v) ~~ {\rm s}^{-1},
\eeq
where $G_0$ is the intensity of the radiation field in Habing units normalized to the typical stellar radiation field, $a_X$ and $b_X$ are the ionization rate coefficients for species $X$ and $A_v$ is the dust extinction. For atomic carbon, $a_C = 2.1 \times 10^{-10}$ s\e and $b_C = 2.6$ \citep[e.g.,][]{mcelroy2013}. 
The reactions associated with FUV are included in the UMIST database.


\subsubsection{X-ray Radiation}\label{sec:xray}

X-ray photons have a significantly longer mean-free-path than FUV photons and serve as an important source of ionization at intermediate column densities \citep{igea99,glassgold07}.  The X-ray ionization rate can be expressed as a function of the column density of hydrogen nuclei above, $N_{\rm Ha}$, and below, $N_{\rm Hb}$, a given location in the disk \citep{igea99,bai09}:
\begin{eqnarray}
\zeta_X &=& L_{X,29} \left( \frac{r}{1\,{\rm au}} \right)^{-2.2} \times \nonumber \\
&& ( \zeta_1 [ e^{-(N_{\rm Ha}/N_{\rm a})^{\alpha_1}} +  e^{-(N_{\rm Hb}/N_{\rm b})^{\alpha_1}}] + \nonumber \\
& & \zeta_2 [ e^{-(N_{\rm Ha}/N_{\rm a})^{\alpha_2}} +  e^{-(N_{\rm Hb}/N_{\rm b})^{\alpha_2}}] ),
\end{eqnarray}
where $L_{x,29} \equiv L_X/10^{29}$ erg s\e, \lx~is the X-ray luminosity,  $\zeta_1 = 6 \times 10^{-12}$\,s\e, $\zeta_2 = 10^{-15}$\,s\e, $\alpha_1 = 0.4$, $\alpha_2 =0.65$, $N_a=1.5\times 10^{21}$\,cm\ee and $N_b = 7 \times 10^{23}$\,cm\ee. Here, we consider only one surface of the disk, adopt an X-ray energy of $T_x=$ 3 keV, and assume the X-ray flux from below is negligible, which is a good approximation as we will show in \S\ref{sec:ionrate}.
The X-ray luminosity correlates with stellar mass, although there is large scatter and a dependence on the stellar evolution model assumed \citep{feigelson05,preibisch05}. For the fiducial model we adopt $L_X = 5 \times 10^{29}$\,erg\,s\e~ \citep[e.g.,][]{bai09}. This is comparable to the X-ray luminosity expected for a 0.3-1 \msun star according to the relation found by
\citet{preibisch05}:
\beq
{\rm log}_{10} (L_X [\rm{erg\, s}^{-1}]) = 30.37 + 1.44 \times {\rm log}_{10} (M/M_\odot), 
\eeq
which was derived empirically from T Tauri sources in the Orion Nebula Cluster. The stellar masses are derived assuming the pre-main sequence tracks from \citet{siess00}.  
Young, active stars that are not accreting also emit X-rays, interestingly, at a rate that is $\sim 2$ times higher than their non-accreting counterparts. We expect this difference occurs because, like cosmic-rays, X-rays are attenuated by the accretion flow. We consider a model with $L_X = 10^{29}$ erg s\e~for comparison. 

\subsubsection{Radionuclides}\label{sec:rn}

In the disk midplane, which is well-shielded from all other ionizing sources, the decay of short-lived radionuclides such as $^{26}$Al, $^{60}$Fe and $^{36}$Cl provide ionization \citep{cleeves13b}. \citet{cleeves13b} find the associated ionization rate depends weakly on time and surface density:
\beq
\zeta_{RN} = 2.5 \times 10^{-19} \left( \frac{1}{2} \right)^{1.04t} \left(\frac{\Sigma}{1\, {\rm g\, cm}^{-2}}\right) ^{0.27} {\rm s}^{-1}.
\eeq
For our {\sc uclchem} calculations, we adopt a constant value of $\zeta = 2.3 \times 10^{-19}$ s\e, equivalent to $t=1$ Myr and $\Sigma = 10$ {\rm g cm}\ee,
throughout the disk. This effectively places a floor on the ionization rate. Ionization produced by short-lived radionuclides dominates only in regions at large radii and close to the mid-plane, where other sources of ionization are absent. 
Long-lived radionuclides also provide ionization, but they contribute at the level of a few $10^{-22}$ s\e\, \citep[e.g.,][]{umebayashi09,padovani2018}.

\begin{figure*}[!ht]
\epsscale{1.1}
\plottwo{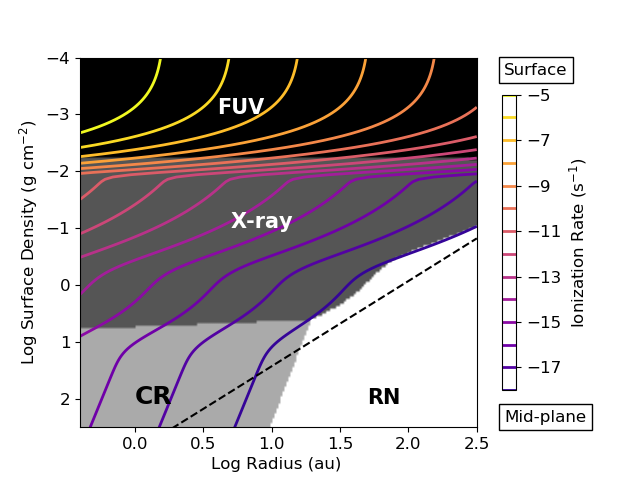}{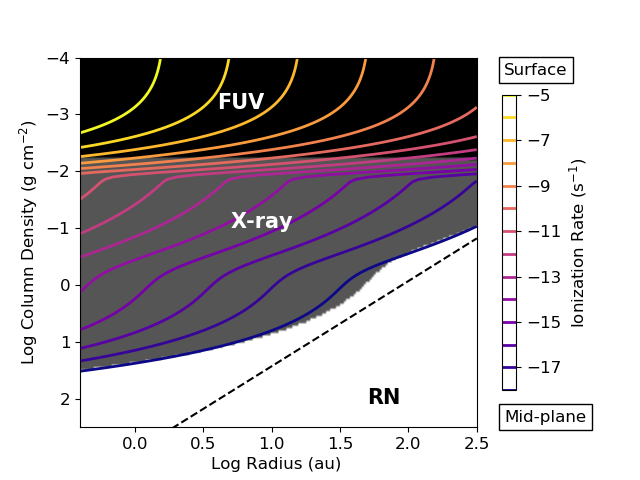}
\plottwo{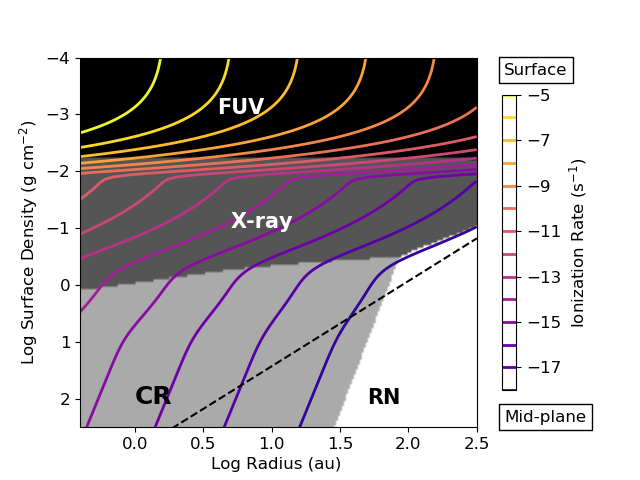}{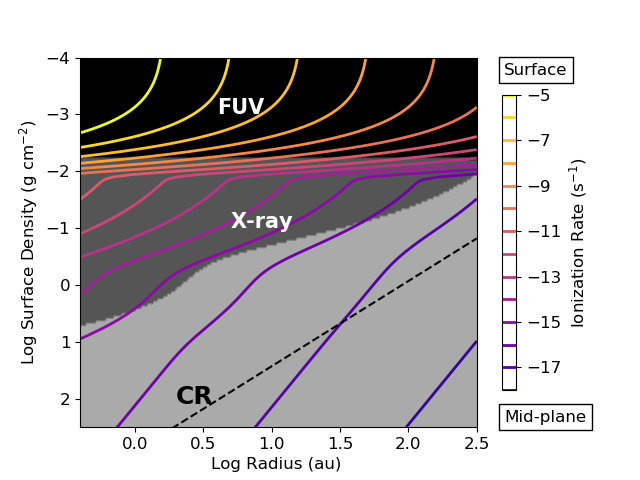}
\caption{Ionization rates as a function of surface density and radius. Shading indicates the dominant source of ionization due to FUV radiation (black), X-ray radiation (dark gray), cosmic rays (CR, light gray) and short-lived radionuclides (RN, white). Top left: Fiducial model for a $1$ \msun~T Tauri star accreting at a rate of $10^{-7}$ \msun\, yr\e (see Table \ref{tab:stars}). Top right: Fiducial model but without CRs. Bottom left: High accretion model ( $\dot M_* =10^{-6}$\,\msun\, yr\e). Bottom right: Model with diffusive CR attenuation in region II ($\zeta \propto r^{-1}$). 
\label{fig:ionization}
}
\end{figure*}

\begin{figure*}[ht]
\epsscale{1.1}
\plottwo{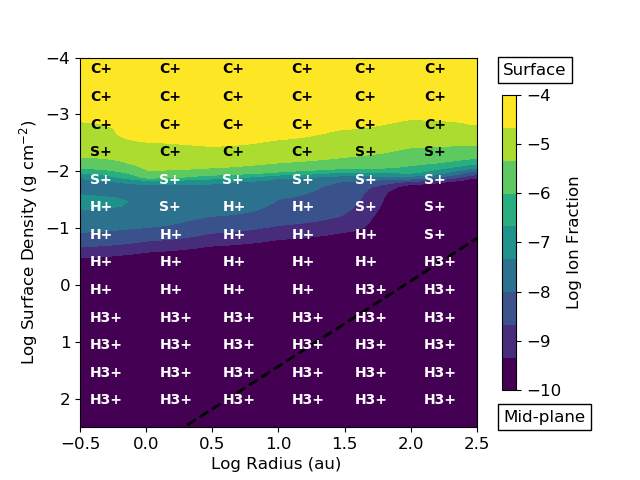}{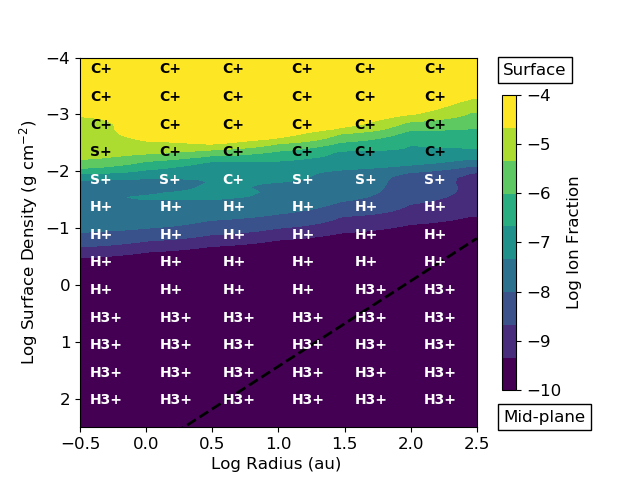}
\plottwo{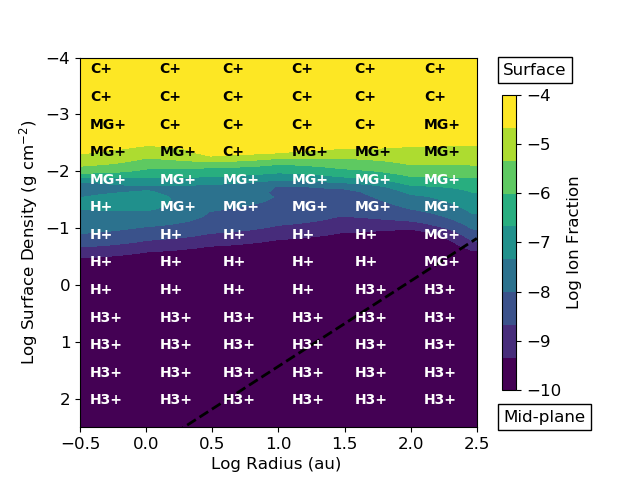}{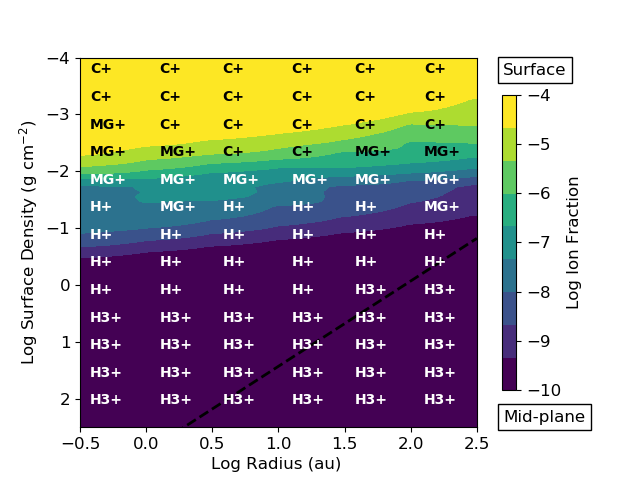}
\caption{ Fraction of positive ions, $n_{\rm ion}/n_{\rm H}$, annotated with the most abundant ion. Top: Fiducial model computed without metals (only H, He, C and O) after $t\sim10^3$ yr of evolution (left) and $\sim 10^5$ yr of evolution (right). Bottom: Fiducial model including Mg, N and Si after $t\sim10^3$ yr of evolution (left) and $\sim 10^5$ yr of evolution (right). 
\label{fig:ionizationfrac}
}
\end{figure*}

\section{Results}\label{results}

\subsection{Ionization Rate Distribution}

Stellar CR feedback produces significant ionization in the disk. Figure \ref{fig:ionization} shows the ionization rate computed using the relations in \S\ref{sec:ionrate} as a function of the disk surface density. The dominant source of ionization is indicated by shading. While non-accreting sources are expected to produce ionization close to the star and throughout a layer at the disk surface, CRs accelerated by accretion shocks extend the degree of ionization near the mid-plane. For our fiducial model (top), the ionization rate exceeds $10^{-17}$ s\e\, for $r \lesssim 10$~au while $\Sigma>10$~g\,cm\ee.  While non-accreting sources may also accelerate CRs \citep[e.g.,][]{rodgers17}, these have a minor affect on the disk ionization, primarily dominating the ionization within a few au. The figure shows that in the absence of locally accelerated CRs, ionization produced by short-lived radionuclides dominates at high-column densities.

The region with high ionizing flux increases with increasing accretion rate. Figure \ref{fig:ionization} shows a comparison between the fiducial and high accretion models for which the accretion rates differ by a factor of 10.  Higher accretion extends the radius at which CR-ionization dominates by a factor of $3$. Although the cosmic-ray flux at the star scales proportionally with the accretion rate, this factor of 10 gain is somewhat offset by the denser accretion flow, which causes additional attenuation. For our fiducial accreting T Tauri stars, ionization caused by CRs is negligible only when $\dot M_*<10^{-9}$ \msun\,yr\e~and the shock becomes too weak.

These results suggest that young, accreting sources may experience higher ionization in their inner disks compared to older, non-accreting sources. This has implications for disk chemistry and angular momentum transport in proto-planetary disks as we discuss below. 

\subsection{Disk Ionization and Chemistry}

The local ionization rate is a crucial factor in disk chemistry \citep{vandishoeck2014}. Thus, enhanced ionization due to CRs may significantly impact the abundance of ionized species, the fraction of free electrons, and the properties of ices. 
Figure \ref{fig:ionizationfrac} shows the ion fraction, the fractional abundance of positive ions, annotated by the dominant ion species throughout the disk. At the disk surface the ionizing flux due to FUV photons dominates and the gas layer is essentially a photo-dissociation region (PDR), where atomic ions such as  C$^+$, S$^+$, H$^+$ and Mg$^+$ are the most abundant. At progressively higher column densities the more abundant atomic ions, such as Carbon, become neutral. The edge of the PDR region occurs around $\Sigma \sim 10^{-2}$\,g\,cm\ee; thereafter H$_3^+$ is the most abundant ion. H$_3^+$ is a well-known product of reactions caused by the CR ionization of H$_2$ \citep{dalgarno2006}: 
\begin{equation}
{\rm CR + H_2 \rightarrow H_2^+ + e^- + CR'}
\end{equation}
\begin{equation}
{\rm H_2^+ + H_2 \rightarrow H_3^+ + H}.
\end{equation}
Consequently, the H$_3^+$ abundance is often used as a direct measure of the CR ionization rate \citep{indriolo2012,gaches19}. However, H$_3^+$ may also be produced by X-ray ionization \citep{gredel01}. In our treatment, both X-rays and CRs result in H$_3^+$ production.  

A variety of additional ion-neutral reactions follow from the presence of H$_3^+$ ions and it serves as a gateway to more complex chemistry \citep{tielens05,dalgarno2006}. Notably, C$_3$H$_5^+$ makes a brief appearance as the most abundant ion in the simple network run at early times with small dust grains. 
This occurs because H$_3^+$ also initiates the formation of hydrocarbon radicals  \citep{tielens05}:
\begin{equation}
{\rm H_3^+ + C \rightarrow CH^+ + H_2}.
\end{equation}
The reaction of CH$^+$ with molecular hydrogen then forms hydrocarbon ions like CH$_3^+$ and CH$_5^+$. Together with C$^+$ hydrocarbon reactions produce increasing molecular complexity.

Like \citet{perez11}, we find that ionized carbon and sulphur are abundant in the disk surface layers due to the stellar FUV field. However, when heavier elements are included in our network, magnesium replaces sulphur as the dominant ion at $\Sigma \sim 10^{-2}$\, g\,cm\ee~and the ion fraction increases by a factor of a few. Reactions with magnesium and nitrogen supplant the formation of complex hydrocarbon radicals close to the disk surface. In general, ionization produced by X-rays and CRs shift the ion/neutral transition region to slightly lower column densities.

\subsection{Time-Dependent Chemical Evolution}

{\sc uclchem} solves for the species abundances as a function of time, which provides insight into the chemical evolution.
Our calculations assume steady-state density and temperature distributions. This is a reasonable approximation on the timescale of the rotational period of the disk:
\begin{equation}
    t_{\rm dyn} \simeq  \frac{2 \pi r}{\sqrt{GM_*/r}} = 1 \left( \frac{M_*}{1\, M_\odot} \right)^{-1/2}\left(\frac{r}{1\, {\rm au}} \right)^{3/2} {\rm yr}.
\end{equation}
Figure \ref{fig:ionizationfrac} shows the ionization fraction at $\sim 10^3$ yr (left) and $10^5$ yr (right). The disk chemistry reaches an equilibrium quickly 
and the abundances at the two times are remarkably similar. The most significant changes occur near the disk surface at large radii. 
There is a brief appearance of C$_3$H$_5^+$ in the PDR transition region (top, left) otherwise the dominant ion species is largely time independent. Since we assume that initially half of the hydrogen is molecular and half is atomic, these changes represent a maximum chemical evolutionary progression for the disk upper layers given time-independent CR and radiative fluxes (see \S\ref{variability} for further discussion). For the remainder of our figures we show results for $t=10^3$ yr. Note that the {\sc uclchem} timestep is set by the shortest reaction rates, such that even at this early time the network has advanced hundreds of iterations.

\subsection{Activation of the Magneto-Rotational Instability}\label{sec:mri}

\subsubsection{Calculation of the MRI-Active Region}

The MRI requires sufficient ionization in order for the gas to couple with the magnetic field. The instability can be characterized by two dimensionless numbers: the magnetic Reynolds number, $Re$, quantifies how well the plasma is coupled with the field, while the Ambipolar diffusion number, $Am$, quantifies how well the neutral gas is coupled to the plasma. The magnetic Reynolds number is defined as:
\beq
Re \equiv \frac{c_s h}{\eta} \simeq 1 \left( \frac{x_e}{10^{-13}} \right) \left( \frac{T}{100~ {\rm K}} \right)^{1/2} \left( \frac{r}{{\rm au}}\right)^{3/2},
\eeq
where $c_s$ is the local sound speed, $h = c_s/\Omega$ is the scale height, $\Omega$ is the Keplerian orbital frequency and $\eta$ is the magnetic diffusivity \citep[e.g.,][]{perez11}:
\beq
\eta = 234 \left( \frac{T}{{\rm K}} \right)^{1/2} x_e^{-1}\,{\rm ~cm}^2\,{\rm s}^{-1},
\eeq
where $x_e = n_e/n_{\rm H}$ is the fractional abundance of electrons by number. 
$Am$ represents a collision timescale between the ions and neutrals:
\beq
Am \equiv \frac{n_{\rm charge}\beta_{\rm in}}{\Omega},
\eeq
where $n_{\rm charge} = x_{\rm charge} n_{\rm tot}$ is the total number density of singly charged species
and $\beta_{\rm in} \simeq 2 \times 10^{-9}$\,cm$^3$\,s\e~ is the collisional rate coefficient for singly charged species to distribute their momentum to neutrals. 

 Numerical simulations suggest the MRI is active for magnetic Reynolds numbers of $10^2-10^4$ \citep{sano02,turner07}. Here, we adopt $Re >3000$, which global disk simulations suggest is the minimum value for fully active MRI \citep{flock12b}.  Early work by \citet{stone98} found the MRI also requires $Am>10^2$; however, more recent simulations including ambipolar diffusion by \citet{bai11} showed that magnetized disks may be MRI unstable for any value of Am. \citet{cleeves13a} note that constraints based on $Am$ also depend on the magnetic field strength and the ratio of thermal to magnetic pressure, $\beta$. 
 Following \citet{cleeves13a} we define the MRI-active region according to $Re >3000$ and $Am>0.1$.


Assuming efficient angular momentum transport via the MRI, the disk mass accretion rate can be written in terms of an $\alpha$-disk model \citep[e.g.,][]{perez11}:
\beq
\dot M = 2 \times 3 \pi \Sigma \nu = 6 \pi \Sigma \times {\rm max}~ \alpha \times \frac{k T}{\mu \Omega},
\eeq
where the effective viscosity is given by $\nu = \alpha c_s h$ and $\mu = 2.4 \times 10^{-24}$ g is the mean molecular weight. The factor of 2 accounts for accretion along the top and bottom disk surfaces. To compute the value of $\alpha$, we adopt the relation between $\alpha$ and $Am$ from numerical simulations \citep{bai11}:
\beq
{\rm max}~ \alpha = \frac{1}{2} \left[ \left( \frac{50}{Am^{1.2}} \right)^2 + \left( \frac{8}{Am^{0.3}} + 1\right)^2  \right].
\eeq

\subsubsection{Cosmic-Ray Mediated MRI}

\begin{figure*}[ht!]
\plottwo{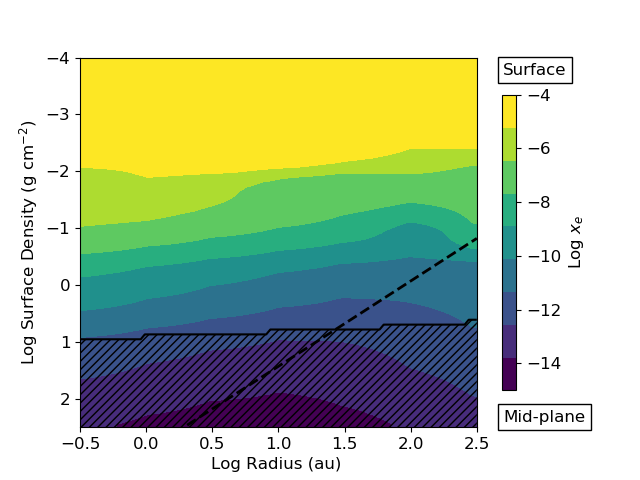}{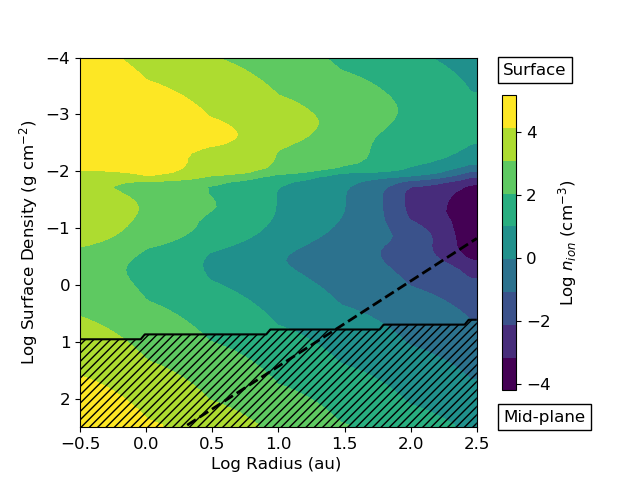}
\plottwo{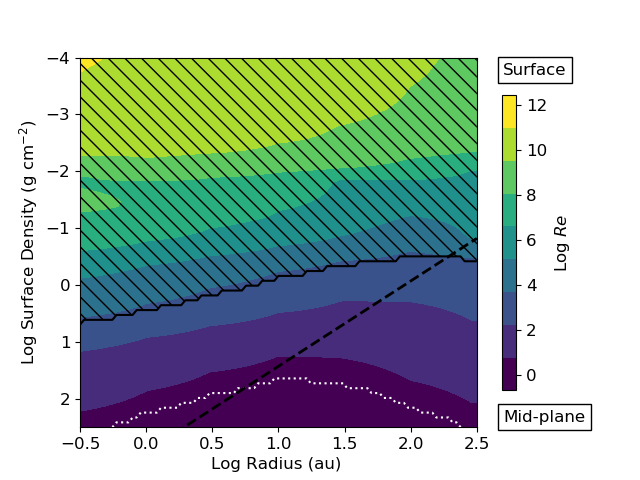}{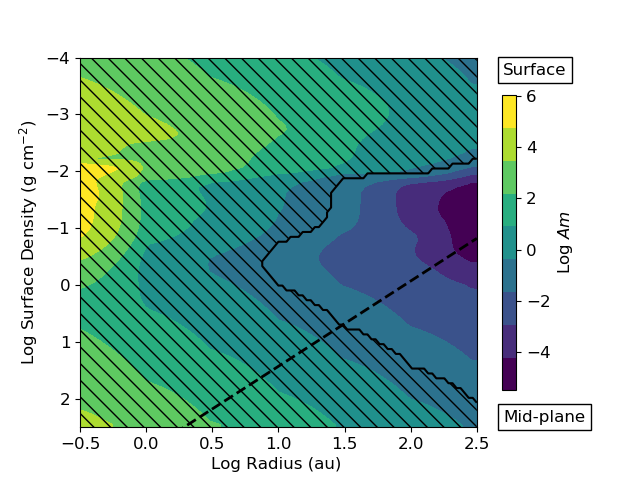}
\plottwo{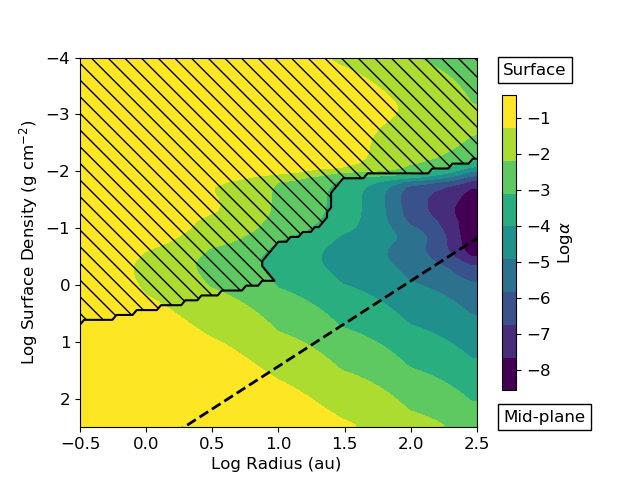}{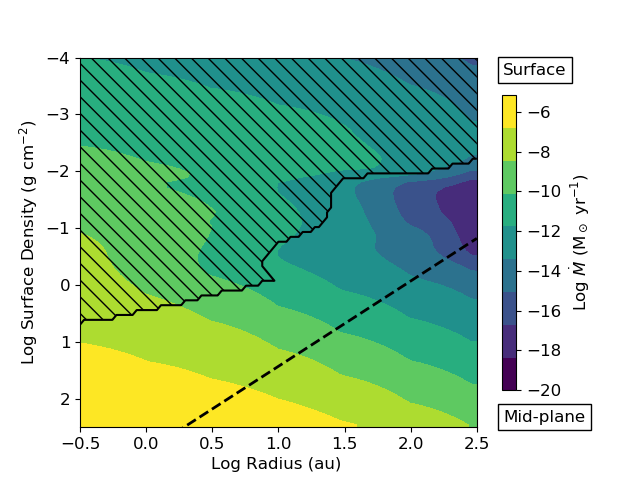}
\caption{ Electron fraction, ion number density, magnetic Reynolds number, Ambipolar diffusion number, viscosity parameter and accretion rate for our fiducial model (1\msun~T Tauri star with accretion rate of $\dot M_* = 10^{-7}$\msun yr\e). The dashed line indicates the surface density at the mid-plane of a MMSN disk. In the top plots, the hashed regions indicates where the CR ionization dominates compared to FUV and X-rays. In the middle plots, the hashed region represents $Re>3000$ (left) and $Am>0.1$ (right), while the white dotted line indicates where $Re=10^2$. In the bottom plots, the hashed region indicates where the MRI is active (i.e., $Re >3000$ and $Am>0.1$). \label{fig:mri}}
\end{figure*}

\begin{figure*}[ht!]
\plottwo{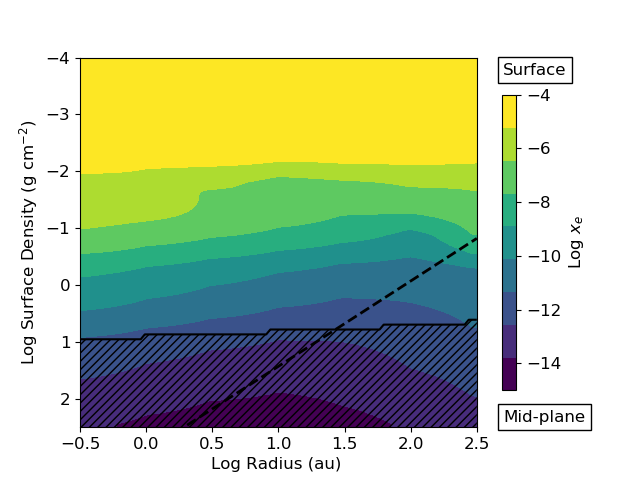}{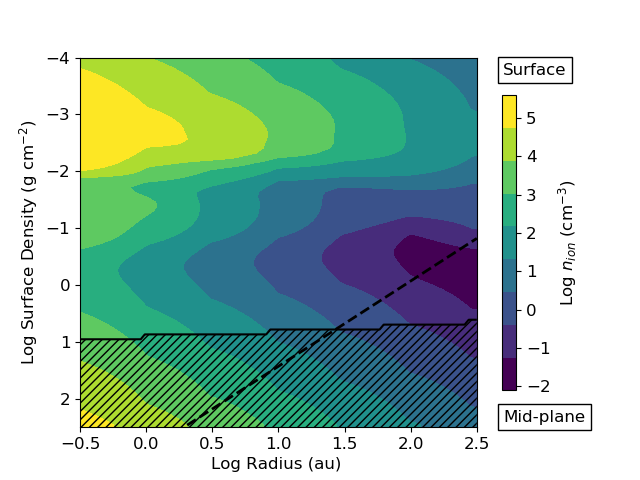}
\plottwo{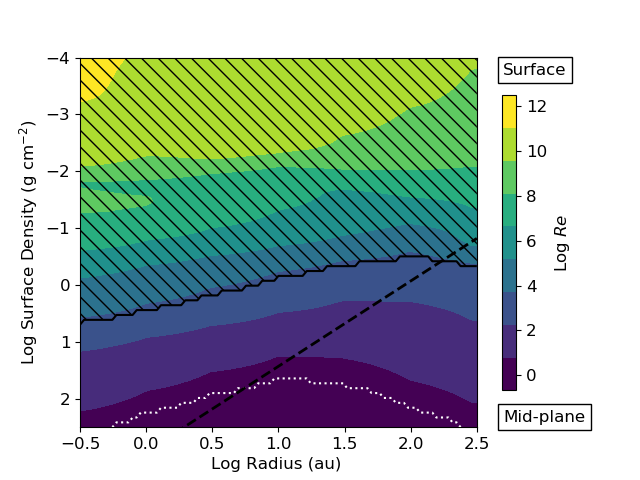}{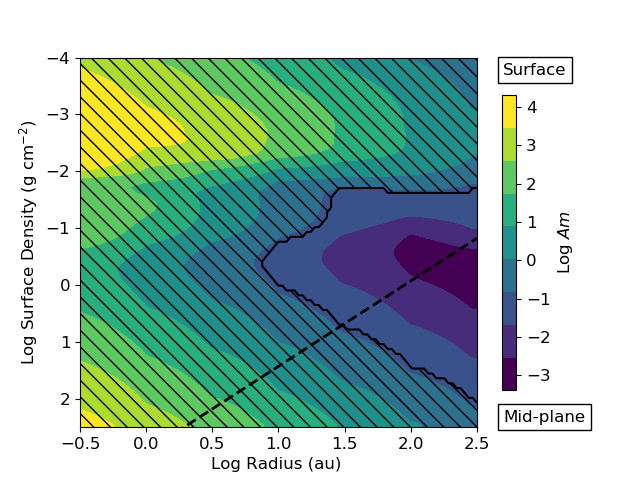}
\plottwo{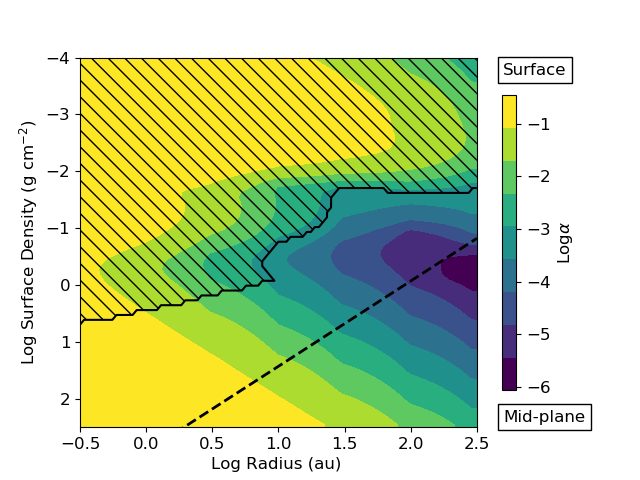}{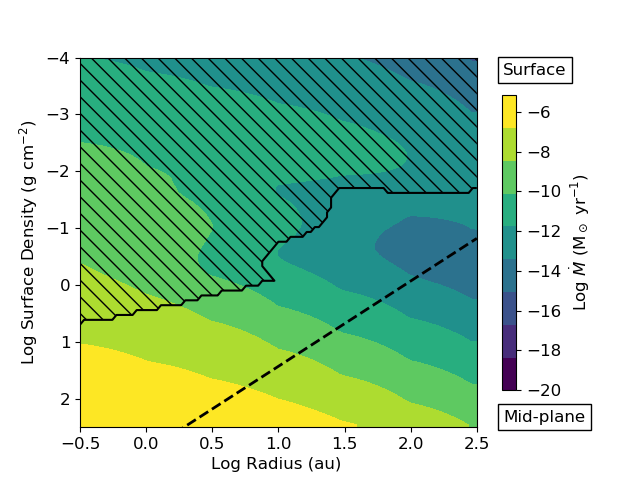}
\caption{Same as Figure \ref{fig:mri} but for the fiducial model with the full network. \label{fig:mri2}}
\end{figure*}

Figures \ref{fig:mri} and \ref{fig:mri2} show the distribution of disk properties for our fiducial model with and without metals. The dashed line in all panels indicates the location of the mid-plane for a minimum-mass solar nebular (MMSN) disk.  The region above this line is representative of older T Tauri type stars, while more massive, circumstellar disks around younger stars would include gas below this line.

The electron fraction (top left panel) decreases smoothly with disk surface density following the declining ionization rate (see also Fig. \ref{fig:ionization}). The number density of positive ions, $n_{\rm ion}$ (top right panel), depends both on the local density, which increases towards the mid-plane, and the local ionization rate. The band in the top third of the plot coincides with the high-ionization FUV-dominated region. A band also appears in the bottom third of the disk. This occurs where the ion number density reaches the minimum value set by the combined ionization from radionuclides and CR. The hashed region indicates where CR ionization dominates over ionization produced by X-ray and FUV photons. 

The middle panels in Figures \ref{fig:mri} and \ref{fig:mri2} show the magnetic Reynolds number and Ambipolar diffusion number. 
As described above, we adopt $Re >3000$ and $Am>0.1$ to determine the MRI-active region.
For the fiducial models, approximately the top half of the parameter space satisfies the magnetic Reynolds criterion, while nearly all satisfies the Ambipolar diffusion criterion.

Altogether, we find that the disk surface layers are MRI-active (bottom right panel),  and  the disk is MRI-active for $\Sigma \sim 1$\,g\,cm\ee~inside $\sim 20$\,au. The MRI active region exhibits accretion rates of $\dot M_* \lesssim 10^{-7}$\,\msun\, yr\e~(bottom left panel). This is consistent with the accretion rate of $10^{-7}$\msun\,yr\e~we adopt for the fiducial model. Although the alpha viscosity in the MRI-active region is reasonably high, $\alpha=10^{-3}-10^{-1}$, suggesting that accretion proceeds efficiently, the low gas density in this region conspires to produce low net accretion. If MRI instability can occur for $10^2 < Re < 3000$, the MRI-active region would extend down to a few $10$\,g\,cm\ee.  Meanwhile, in all scenarios, the disk mid-plane, $\Sigma \lesssim 10^2$\,g cm\ee, is a dead-zone.

We find the addition of N, Mg, Cl, and Si to the chemical network have a minor impact on the disk properties. The magnetic Reynolds number and alpha viscosity increase slightly throughout the disk, however the boundaries of the MRI active region remain similar.

 The dust grain size has a significant effect on the MRI, which has previously been noted in prior work \citep{sano00,wardle07,glassgold17}. This result remains true in the presence of high CR fluxes. The left panel of Figure \ref{fig:mricomp} shows the MRI-active region 
for the fiducial model and the models with small ($a_{\rm gr}= 0.1\, \mu$m) and large  ($a_{\rm gr}= 10\, \mu$m) dust grains. By efficiently binding electrons, small grains reduce the MRI-active region by an order of magnitude. For large grains the active region reaches surface densities of $\Sigma \sim 10$ g cm\eee~ and most of the outer disk becomes MRI-active. 

We plot the boundary of the MRI-active region for a variety of models, including those shown in Figure \ref{fig:mri} and \ref{fig:mri2}, in Figure \ref{fig:mricomp}. We discuss the impact of other parameters on the MRI-active region below.

\subsection{Cosmic Ray Attenuation and Magnetic Turbulence}\label{cratten}

\begin{figure*}[ht]
\epsscale{1.1}
\plotone{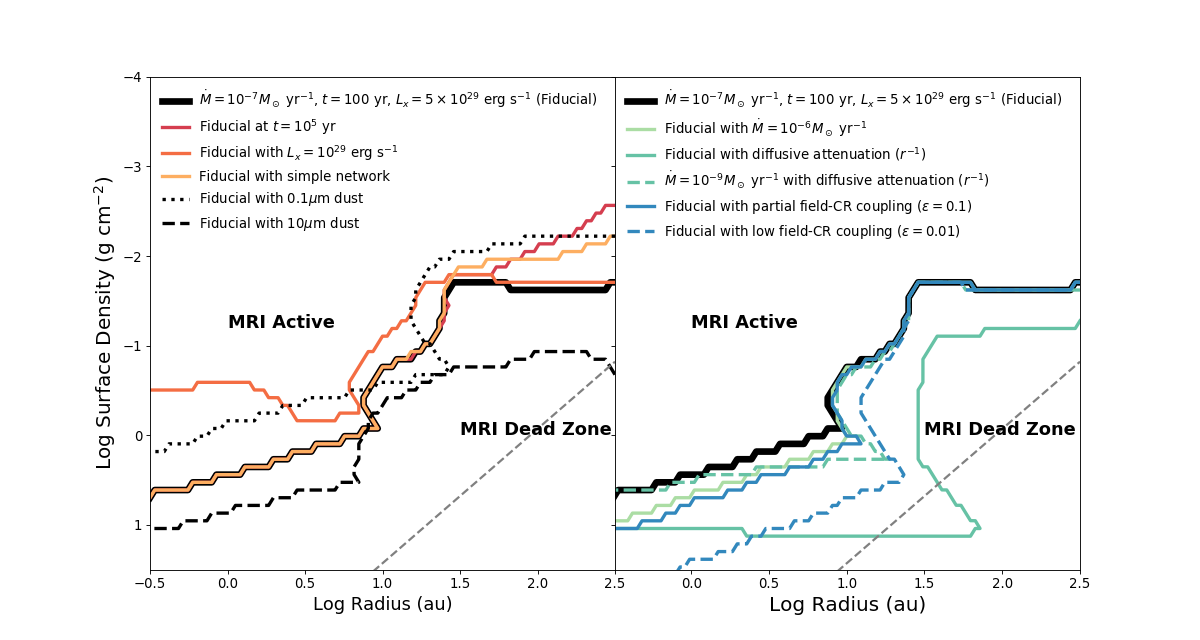}
\caption{ Boundary between the MRI active ($Re>3000$, $Am>0.1$) and MRI dead zones for the fiducial model (thick black line) and variations. \label{fig:mricomp}}
\end{figure*}

The coupling between the mean magnetic field and the CRs is mediated by the degree of turbulence \citep{rodgers17,bai19,thomas19}. Within dense cores, the level of turbulence depends on a variety of complex variables including the mean magnetic field and initial conditions inherited from the environment, infall from the gas envelope and local turbulent driving by the protostellar outflow and disk wind 
\citep{offner17}. CRs, by themselves, may also excite magnetic waves, thereby driving turbulence \citep{schlickeiser2016}. Consequently, the number of scatterings and how the CRs diffuse change significantly \citep{rodgers17}. 
To investigate this uncertainty, we vary the radial attenuation between the free-streaming and diffusive limits. 
In the fiducial models we assume that CRs free-stream once they leave the accretion flow (i.e., $a=2$). Figure \ref{fig:mricomp} shows the change in the MRI-active region if $a=1$ for the fiducial and low-accretion models. In these model, the CR ionization rate exceeds $10^{-14}$\,s\e~at the disk surface for $r \lesssim 30$\,au. This higher ionization extends the MRI-active region to surface densities of a few g\,cm\ee~for these radii. This suggests that only extreme cosmic ray fluxes can trigger the MRI in regions near the midplane for radii greater than $\sim30$\,au.

The fiducial model also assumes the CRs are well-coupled to the gas until they reach the disk inner edge. This means that the CRs must travel upstream through the accretion flow and thus attenuate significantly before reaching the disk. If the accretion flow is turbulent or if it is variable, such that the column density of gas along the CR path is reduced, the CRs will experience less attenuation. We introduce a parameter $\epsilon$ to quantify the interaction between the CRs and the accretion flow. A value of $\epsilon=1$ indicates the accretion rate is constant and the CRs traverse the full accretion column,  i.e., they are perfectly coupled to the magnetic field in the accretion column and thus suffer energy losses due to their interactions with the gas as they travel between the star and the disk.  Lower values of $\epsilon$ represent imperfect coupling with the magnetic field and/or a decline in the upstream accretion rate by a factor of $\epsilon$,  two factors that reduce the magnitude of the CR energy losses. If the CRs are able to diffuse out of the high-density accretion column, then they will propagate through significantly less material. So $\epsilon=0.1$ represents a scenario in which CRs on average traverse only 10\% of the accretion column gas. Alternatively, the CRs may interact with a smaller column of gas if accretion has declined by the time the CRs escape from the accretion shock, and the gas density in the accretion column is lower. Thus, $\epsilon=0.1$ may also represent the case when the accretion has declined by a factor of 10, but the CRs remain coupled to the magnetic field. Consequently, $\epsilon$ provides a crude way of parameterizing some of our uncertainty about the energy losses experienced by the CRs between the star and the disk.

The right panel of Figure \ref{fig:mricomp} shows the MRI-active region for $\epsilon=0.1$ and $\epsilon=0.01$.  Figure \ref{fig:mricomp} shows that reducing the attenuation caused by accreting gas increases the MRI-active region to higher surface densities as expected. However, the mid-plane of the disk remains MRI-inactive. 

\section{Discussion}\label{discussion}

\subsection{Implications for Disk Evolution: Accretion Variability}\label{variability}

The luminosity of young stellar objects is known to be variable on timescales ranging from hours to hundreds or possibly thousands of years \citep{audard14}. 
The magnitude in the change in luminosity appears to correlate with frequency in that order unity changes are common, while changes by orders of magnitude, like FU Ori-type bursts, are rare \citep{offner2011,hillenbrand15}. Short term luminosity changes lasting months to a few years are generally attributed to inner disk variation, which could be caused by changes in disk morphology or the accretion flow \citep[e.g.,][]{rebull14}.

Our results show CRs accelerated by accretion shocks impact disk chemistry and thereby mediate stellar accretion. Consequently, CR-feedback provides a mechanism to self-regulate accretion as follows. If the accretion rate declines, stellar CRs are less attenuated by the accretion flow, and the CR flux reaching the disk increases. This causes the MRI-active region to expand, thereby increasing the disk accretion rate. Once the accretion rate increases, the CR flux reaching the disk declines, reducing the MRI active region and so on. 
 We expect CR-related accretion variation to occur on time scales of a few years, as dictated by the disk dynamical time at the edge of the MRI-active region (e.g., Figure \ref{fig:mricomp}). However, the magnitude of variation depends on the details of the CR propagation.  

Our models predict that CRs will be efficiently accelerated down to accretion rates of $\sim 10^{-9}$\,\msun\,yr\e. Thus, CR-feedback is likely important for most T Tauri stars. Higher accretion rates produce higher CR fluxes, however, the attenuation of the dense accretion flow is also more severe. This suggests that CRs accelerated in protostellar accretion shocks may not influence the chemistry of their disks as strongly. However, protostellar outflow may advect some of these CRs outward \citep[e.g.,][]{rodgers17} and thereby allow them to influence the chemistry of the natal core and molecular cloud. 

 Constraining the luminosity variability of deeply embedded protostellar sources is challenging, since it is difficult to directly observe bursts. However,  observations of molecules in the gas phase, like CO, which quickly freeze out at high densities and cold temperatures, are thought to provide evidence of past accretion bursts \citep[e.g.,][]{jorgensen2015,frimann2016}. Although there are a number of uncertainties, this line of inquiry suggests that accretion variability of a factor of a few to 10 is relatively common in young objects on thousand-year time scales. The origin of this variability is debated, and it is often attributed to perturbations by a companion or gravitational disk instability rather than MRI variation \citep{audard14}. However, independent of origin, the byproduct of the enhanced accretion, CRs, are likely to also impact the observed abundances. In general, we expect CRs to enhance the gas phase CO and thus reduce the luminosity needed to produce the observed CO emission \citep[e.g.,][]{gaches19}. Thus, CRs accelerated via accretion shocks have implications not only for modulating disk variability but for constraining the true frequency of accretion bursts and their underlying physical origin.



\subsection{Implications for Complex Disk Chemistry}

The additional ionization provided by local CRs also has implications for the formation and abundance of complex organics. For example, ionization is important to the gas-phase production of methyl-cyanide (CH$_3$CN), since its main gas-phase formation pathway requires CH$_3^+$ (HCN + CH$_3^+$) \citep{walsh14,oberg15}. Ionization rates of $10^{-14}$\,s\e~could explain the large ratio of CH$_3$CN to HCN observed in the MWC 480 protoplanetary disk \citep{oberg15}.  We find the ratio, CH$_3$CN/HCN $> 0.1$, and large CH$_3$CN abundance are reproduced in our model through gas phase chemistry driven by ionization for $\Sigma < 10^{-2}$\,g\,cm\ee~ at 100\,au. However, disk turbulence that produces vertical mixing may still be important if CH$_3$CN is abundant outside of the region in which it is produced, as predicted by \citet{oberg15}.


\subsection{Implications for Disk Turbulence}\label{turb}

Despite the enhanced ionization rates at the disk surface, our models predict the MRI is weak or absent at the disk midplane. This is consistent with recent observations of protoplanetary disks using the linewidth of CO isotopologues to place constraints on disk turbulence. The level of turbulence observed in protoplanetary disks is quite low and recent observations place upper limits of $v_{\rm turb} \lesssim 0.08\,c_s$ on the turbulence in the outer disk ($r\gtrsim $\,50\,au) near the midplane \citep{flaherty15,flaherty17,flaherty18}. 
Even with ALMA, resolved observations of protoplanetary disks are challenging, however, and there are many uncertainties, e.g., \citet{teague16} report $v_{\rm turb}= 0.2 - 0.4c_s$ for TW Hya.

\subsection{Comparison with Past Work}\label{pastwork}

Our results are consistent with \citet{bai09}, who adopt a simpler network but include heating and cooling. For the same X-ray luminosity, $L_X=5 \times 10^{29}$\,erg\,s\e, but no CR ionization they predicted an MRI-active region down to $\Sigma=30-50$\,g\,cm\ee, including the MMSN disk midplane beyond 50\,au,  given an MRI-active threshold of $Re>100$. \citet{bai09} showed that including a constant CR ionization rate of $\zeta = 10^{-17}$\,s\e~further extends the MRI-active region. We find a similar result for the same magnetic Reynolds number threshold modulo our higher local CR ionization rate (see dotted line on our Figures \ref{fig:mri} and \ref{fig:mri2}). More recent numerical work has since revised the magnetic Reynolds number threshold for active MRI \citep{flock12a}, which directly leads to the conclusion that the midplane is an MRI dead zone.

Our models produce a similar ionization rate distribution as \citet{cleeves13a} given similar X-ray, FUV, radionuclide and CR parameters. However, we find electron abundances that are lower by a factor of 10-100 near the mid-plane. This difference leads \citet{cleeves13a} to conclude that under similar conditions most of the disk is MRI-active (e.g., their Figure 14), while we conclude the midplane generally inactive. They adopt a more accurate treatment of X-ray ionization, including a hard X-ray spectrum, however the discrepancies in our results may also be due to differences between our chemical networks. The network adopted by \citet{cleeves13a} is simpler and includes fewer grain reactions. Recent observations suggest minimal MRI activity at large radii near the disk midplane, which we find here ( see \S\ref{turb}).


\begin{figure}
\plotone{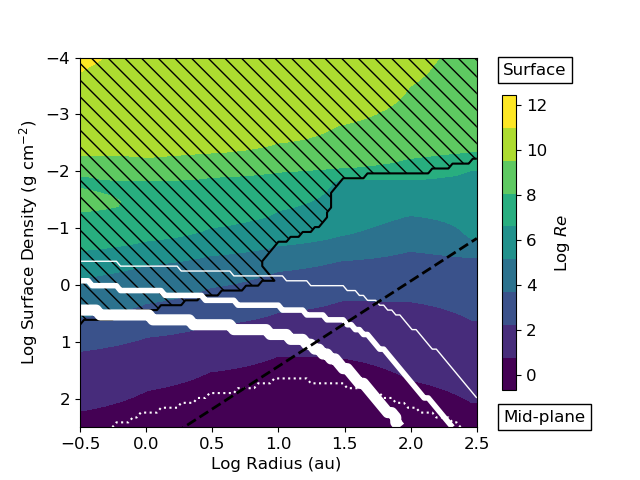}
\caption{ Magnetic Reynolds number for the fiducial model with the corresponding MRI-active region over-laid. The white solid lines indicate the MRI limit in the case of anti-aligned rotation and magnetic field vectors for $B_z=1$\,G (thick), $B_z=$0.1\,G (intermediate) and $B_z=$0.01\,G (thin) as defined by Equation \ref{eq:hall}. \label{fig:hall}}
\end{figure}

\subsection{Impact of the Hall Effect on Disk Stability}

In our analysis we assume the diffusivity is dictated by ohmic dissipation, a regime in which both ions and electrons are decoupled from the magnetic field, following the convention of a number of past studies \citep[e.g.,][]{stone98,bai09,perez11,cleeves13a}.  
However, electrons and ions tend to be well-coupled to the field at lower density. Ambipolar diffusion occurs when both ions and electrons are coupled to the field. This effect dominates at low column densities, when the gas is well-ionized, namely at the disk surface and in the outer disk. Hall diffusion, in which only electrons are coupled to the field, is important at intermediate densities, while ohmic diffusion only  dominates at high densities \citep{wardle07}. In this work we are mainly concerned with the impact of CRs on the MRI-active region at intermediate to high densities and, thus, the distinction between the ohmic and Hall diffusion-dominated regions is potentially important. The Hall effect has the peculiar feature of depending on both the magnitude and direction of the magnetic field. \citet{wardle12} compare the relative importance of Hall and ohmic diffusion on the MRI and demonstrate that Hall diffusion can increase or decrease the MRI-active region by an order of magnitude depending on the field orientation with respect to the rotation axis of the disk. Reproducing their analysis here is beyond the scope of this work; however, we can assess the limit in the Hall-dominated diffusion regime for $B_z < 0$ (magnetic field anti-aligned with the rotation axis) in which the Hall effect acts to stabilize the disk to the MRI.  In this regime, \citet{wardle12} derive a criterion for stability based only on the magnetic field strength and the fractional ionization \citep{wardle12}:
\begin{equation}
   x_e \lesssim \frac{1}{2}\frac{1.4 m_H c \Omega}{eB} \simeq \frac{1.5 \times 10^{-11}}{B / {\rm G}} \frac{(M/M_\odot)^{1/2}}{(r/{\rm au})^{3/2}}. \label{eq:hall}
\end{equation}
Figure \ref{fig:hall} shows the MRI-boundary for our fiducial model with the limit imposed by the Hall effect from Equation \ref{eq:hall} for three different field strengths. We overlay this on the magnetic Reynolds number, which is defined assuming ohmic dissipation in our calculations.  This figure shows the MRI will be suppressed below the critical column density our models predict, but only in the inner disk ($\lesssim$ a few au) and only for weak fields. In contrast, for fields aligned with the disk rotation axis, $B_z>0$, \citet{wardle12} show the Hall effect enhances the MRI. The Hall effect has also been studied directly in a small number of magnetohydrodynamic simulations. There the Hall-shear instability is found in the $B \cdot \Omega>0$ case \citep{kunz08} and is observed  to produce a substantial amount of radial transport of angular momentum even in the ``dead zone"  \citep{lesur14, bai14, simon15, baistone17}. 
Thus, in this limit we expect consideration of the Hall effect will increase the MRI-active region beyond that predicted by our models. However, the nature of the field configuration in accretion and protoplanetary disks remains poorly constrained both observationally and theoretically. Additional constraints on the true magnetic field strength and orientation in disks are required to make stronger claims about the influence of the Hall effect on the MRI.

\subsection{Caveats}

We assume that photons and CRs enter the disk vertically through its top surface and ignore contributions from photons and CRs that enter the disk ``side-ways" at the inner radius. Consequently, our results underestimate the ionization at the midplane in the inner $\sim10$\,au due to both X-rays and CRs \citep{igea99,rodgers17}. This additional ionization would likely extend the MRI-active region throughout most of the vertical extent within a few au of the source. In addition, high temperatures, which exceed $\sim 800$ K in the very inner disk, will thermally ionize alkali atoms, yielding an ionization fraction of up to $10^{-8}$ s$^{-1}$ \citep[e.g.,][]{fromang02,desch15}.

While we account for all the main sources of ionization, our estimation of the MRI is relatively simplistic. We do not directly model the fluid dynamics of the disk \citep[e.g.,][]{simon18}. Thus, we may miss other dynamical effects. For example, \citet{gole18} find that the FUV-irradiated surface layers may affect the magnetic structure of the disk by transporting magnetic flux to the mid-plane, thereby suppressing the MRI. 

 We do not self-consistently model grain growth, which is 
especially important at the disk midplane, where grains colagulate and sediment
\citep{dullemond04}. A vertically variable dust distribution may modify the details of our results somewhat. 

We also do not explicitly follow charged grains. Charged grains may impact our MRI analysis via the Ambipolar diffusion number, where the number density of charged species, $n_{\rm charge}$, can be written including positively and negatively charged grains, i.e., $n_{\rm charge} = n_{\rm ion} + n_{\rm Q+} + n_{\rm Q-}$ \cite[e.g.,][]{perez11}. To roughly constrain the impact of charged grains on our results we use the abundance of electrons sticking to dust grains to compute the average charge per grain. We find that grains near the disk surface ($\Sigma \lesssim 0.1$ g cm\ee) are on average strongly positively charged. At higher surface densities ($\Sigma > 0.1$ g cm\ee) grains become slightly negatively charged. Including the number density of charged grains in the calculation of $Am$ has a small effect, since it mainly enhances Am near the disk surface, which is already MRI active. A more sophisticated grain treatment, including polycyclic aromatic hydrocarbons (PAHs) is required to quantify the impact of charged grains on the MRI with more accuracy, although the impact of small grains is expected to be modest \citep{bai11b,xu16}.

We also make simple assumptions about the geometry of the magnetic field, which has an outsize impact on the propagation of the CRs and local ionization rate, both by influencing the degree of attenuation and the severity of the energy losses \citep{fraschetti18}. Some of the stellar CRs may be directed away from the disk by the outflow \citep{rodgers17}, which may cause CR beaming and further reduce the CR flux reaching the disk. Full magneto-hydrodynamic simulations, including self-consistent modelling of the stellar, disk and outflow fields together with cosmic ray propagation, are required to study the impact of realistic magnetic field configurations \citep{romanova15,bai19}. 

\section{Conclusions}\label{conclusions}

In this work, we use the gas-grain astrochemistry code {\sc uclchem} to investigate the impact of cosmic rays accelerated in accretion shocks on the chemistry of circumstellar disks. We include all major sources of ionization, namely FUV radiation, X-rays, CRs and radionuclides, and model the attenuation of each as a function of radius and gas surface density.  
We explore accretion rates of $\dot M_* \sim 10^{-9}-10^{-6}$\,\msun\,yr\e~and assess how different treatments of CR propagation and magnetic field-CR coupling influence the ionization rate. We show that locally accelerated cosmic rays can produce ionization rates $\zeta \gtrsim 10^{-14}$\,s\e~at the disk surface within 10\,au of the central star. Our results show that local shock-accelerated CRs can produce ionization rates that are several orders of magnitude higher than the rate associated with the Galactic cosmic-ray background. 

The cosmic-ray flux from the star enhances the disk ionization at intermediate to high surface densities ($\Sigma > 10$\,g\,cm\ee) particularly within 10 au of the star. We show C$^+$, S$^+$ and Mg$^+$ are the dominant ions in the disk surface layers, while H$_3^+$ ions dominate at surface densities above 1.0\,g\,cm\ee. 

We use the electron fraction and ion number density to predict the radii and surface densities where the magneto-rotational instability (MRI) is active. We show that ionization by local cosmic rays extends the MRI-active region towards the disk mid-plane. However, the mid-plane of a minimum mass solar nebula disk remains an MRI-dead zone unless cosmic-rays propagate outwards through diffusion ($\zeta \propto r^{-1}$).  

We note an important feedback loop circumscribed by the relationship between the cosmic-ray flux, which scales with the accretion rate, the column density of the accretion flow, which attenuates cosmic rays, and the MRI, which is enhanced by  cosmic-ray ionization and in turn promotes accretion. This relationship limits the efficacy of cosmic-ray ionization in disks at very high accretion rates, i.e., in protostellar disks. In contrast, the disks around T Tauri stars experiencing accretion rates of $\dot M_*  = 10^{-9}-10^{-7}$\,\msun\,yr\e~ will be significantly influenced by cosmic rays accelerated by accretion.

In summary, we conclude that cosmic-rays are a substantial source of ionization in the disks of young, accreting stars. Future work is required to explore the impact of magnetic field geometry on the cosmic ray attenuation and spectrum evolution.

\software{ {\sc uclchem} \citep{holdship17}, {\sc matplotlib} \citep{matplotlib2007}, {\sc NumPy} \citep{numpy2011}, {\sc SciPy} \citep{scipy2001},  {\sc JupyterLab}}.

\acknowledgments
SSRO and BALG acknowledge support from the National Science Foundation (NSF) grant AST-1510021. SSRO was also supported by NSF CAREER grant AST-1650486. 
The authors thank an anonymous referee for helpful comments that improved the manuscript. The authors also acknowledge helpful discussions with Xuening Bai and Marco Padovani.
The authors acknowledge the Texas Advanced Computing Center (TACC) at The University of Texas at Austin for providing HPC resources that have enabled the research reported within this paper.

\bibliography{ref.bib}
\end{document}